\def\ds{D_{\rm s}}

\def\Dcut{ D_{\rm cut-off}}
\def\sol{ {\rm M}_\odot\, {\rm kpc}^{-3} }
\def\msun{ {\rm M}_\odot }
\def\kpc{ {\rm kpc} }
\def\kms{ {\rm km} {\rm s}^{-1}  }
\def\spose#1{\hbox to 0pt{#1\hss}}
\def\lta{\mathrel{\spose{\lower 3pt\hbox{$\sim$}}
    \raise 2.0pt\hbox{$<$}}}
\def\gta{\mathrel{\spose{\lower 3pt\hbox{$\sim$}}
    \raise 2.0pt\hbox{$>$}}}

\documentstyle[12pt,aasms]{article}

\begin{document}

\title{Microlensing Maps for the Galactic Bulge}
\author{N.W. Evans}
\affil{Room 2-367, Department of Mathematics, Massachusetts Institute
of Technology, Cambridge, MA 02139, USA}

\begin{abstract}

Microlensing maps -- that is, contours of equal numbers of events per
$10^6$ source stars -- are provided for the inner Galaxy under two
alternative hypotheses : (1) the bulge is an oblate axisymmetric spheroid
or (2) the bulge is a prolate bar. Oblate spheroids yield a total of
$\sim 12$ events per year per $10^6$ stars at Baade's Window ($\sim
15$ events if the disk is maximal). The event rate is slightly lower for
prolate bars viewed at $\sim 45^\circ$ and the maps have a characteristic
asymmetry between positive and negative longitudes. Prolate bars can yield
mild amplifications of the event rate if viewed almost down the
long axis. The disk provides the dominant lensing population on the
bulge major axis for $|\ell | \gta 6^\circ$. Measurements of the rate
at major axis windows can test for disk dark matter or maximal disk models.

\end{abstract}


\section{INTRODUCTION}

The startling reports of enhanced rates of microlensing towards the
Galactic bulge (Alcock et al. 1994, Udalski et al. 1994a, Udalski et
al. 1994b) confront us with immediate challenges. Are the observations
consistent with the known populations of deflectors? Do the microlenses lie
in the Galactic disk or in the bulge or even -- as recently suggested by
Paczy\'nski et al. (1994) -- in a bar? The aim of this Letter is to
suggest observational tests that can provide answers to these questions.

\section{BULGE, BAR AND THICK DISC MODELS}

Let us first introduce models for the components of the inner Galaxy.
A simple, though idealised, oblate axisymmetric bulge model is provided by
Evans \& de Zeeuw (1994, hereafter EZ). Their starting point is Becklin
\& Neugebauer's (1968) observation that the emissivity profile of the inner
bulge is scale--free and falls like $\rho \sim r^{-1.8}$. So, a good
representation of the density profile of the cusp is provided by a
scale--free oblate power--law model (Evans 1994, hereafter E94). The
density of the model is
\begin{equation}
\rho \sim 5.9 \times 10^8
  {R^2 + 0.87 z^2 \over (R^2 + 1.23 z^2)^{1.9}}
 \,\,\sol
\end{equation}
where $R$ is the cylindrical polar radius and $z$ the height above the
galactic plane in kiloparsec. The model has an apparent axis ratio
$a_2/a_1$ of $0.75$ (Habing 1988) and a line of sight velocity
dispersion at Baade's Window of $119\, \kms$ (c.f. Kent 1992). Outside of
the inner kiloparsec, the luminosity density of the galactic bulge falls
like $r^{-3.5}$. We do not model this in detail, but truncate (1) at some
distance $\Dcut$, beyond which the density is assumed to vanish. We choose
$\Dcut$ to be $3\,\kpc$ so that the total bulge mass is $\sim 1.9 \times
10^{10}\, \msun$. The advantage of EZ's bulge over Kent's (1992)
hydrodynamical model is that the self--consistent distribution of proper
motions for the former is explicitly available (see EZ, Appendix B).

An alternative viewpoint is to regard the bulge as a prolate
bar viewed somewhat broad--side on (de Vaucouleurs 1964). A number of
investigators have concluded that the evidence from neutral and ionised gas
motions (Binney et al. 1991, Blitz \& Spergel 1991) and star--counts (Weinberg
1992, Stanek et al. 1994) suggests that the Galaxy is barred. We shall use a
prolate power--law model (E94) to represent the bar (c.f., Binney et al. 1991)
\begin{equation}
\rho \sim 3.7 \times 10^8
  {x^2 + 0.89 (y^2 +z^2) \over (0.67x^2 + y^2 + z^2)^{1.9}}
 \,\,\sol
\end{equation}
Here, the  $x$--axis defines the long axis of the bar, which is oriented
at an angle $\theta$ to the line joining the Sun to the Galactic Center
and the model is truncated at $\Dcut =3\,{\rm kpc}$. Our bulge (1)
and bar (2) models have the same total mass and roughly the same apparent
flattening. Although all investigators agree that the near--side of the bar
lies at positive Galactic longitudes ($\ell >0$), there is no concensus
over the angle $\theta$ at which we view the major axis. Note that  EZ's
distribution of proper motions for the power--law models assume the figure
is fixed in inertial space. This is a fair approximation if the pattern
speed of the bar is small (c.f. Blitz \& Spergel 1991).

Lastly, we need a model of the Galactic disk, whose density we assume
to be exponentially declining on spheroidal surfaces
\begin{equation}
\rho \sim 3.9 \times 10^8 \exp( -0.29\sqrt{R^2 + 144 z^2} )\,\,
\sol.
\end{equation}
The axis ratio $a_2/a_1$ is $0.083$, as suggested by Bahcall \& Soneira's
(1980) analysis. The scale--length of the disk is $3.5\,\kpc$. The
normalisation is chosen to recover the local column density of
$\sim 50\,\msun \,{\rm pc}^{-2}$ (Gould 1990). It is sometimes suggested
that this data may be misleadingly low. An upper limit is provided by
`maximal disk' models, in which disk matter provides almost all
the local centrifugal balance (e.g., Alcock et al. 1994). If
the disk is maximal, then the density (3) is roughly doubled.
Self--consistent velocity distributions for thick disks are not known
-- but this is not too serious as the random motions are much smaller
than the systemic rotation of $\sim 200\, \kms$.

\section{MICROLENSING MAPS}

The contributions of different deflecting populations can be distinguished
by the variation in microlensing rate as a function of Galactic longitude and
latitude. A simple way to picture this is provided by {\it microlensing maps}
-- that is, contours of equal numbers of events per $10^6$ source stars.
The maps are useful because they are lots of windows of low extinction
towards the bulge. Even along heavily obscured lines of sight, the main
effect of interstellar extinction is to reduce the number of detectable
source stars. This just alters the overall normalisation of the rate.

The microlensing rate $\Gamma$ is the reciprocal of the time between events
averaged over all possible lensing configurations. So, $\Gamma$ depends
not only on the densities but also the distributions of masses and proper
motions of the sources and the deflectors. Kiraga \& Paczy\'nski (1994,
hereafter KP) introduced a refinement by modulating the density of the
sources by a factor of $\ds^{2\beta}$, where $\ds$ is the distance
of the source from the sun. This is a crude way of modelling the tendency to
observe fewer stars at the far side of the bulge. The value $\beta = -1$
is reasonably close to what is observed (Depoy et al. 1993). To construct
microlensing maps, the rate is calculated at any Galactic longitude and
latitude ($\ell,b$) using eq. (11) of KP. The tangential velocity components
of the bulge or bar stars are randomly drawn from the distributions of
proper motions of the power--law models (EZ). The distribution of stellar
masses is taken as a Gaussian in $\log_{10} M$ for masses greater than
$0.35\, \msun$ and flat below $0.35\, \msun$ to a lower cut-off of
$0.01\, \msun$ (Kroupa, Tout \& Gilmore 1990).

Figure 1 shows a map for the bulge (1) and disk (3) model of the inner
Galaxy. The contributions from the different deflector populations have
been separated. The full (broken) contours delineate the number of events per
year per $10^6$ source stars caused by deflectors in the bulge (disk). If the
disk is maximal, then the share from the disk lenses is nearly doubled. The
optical depth at Baade's Window $(\ell = 1.0^\circ, b= -3.9^\circ)$ is
$6.3 \times 10^{-7}$ for microlensing by bulge lenses, $5.0 \times 10^{-7}$
for disk lenses. The frequency of events is $\sim 12$ per year per $10^6$
stars, of which $\sim 70 \%$ are caused by bulge lenses (c.f. KP). This rises
to $\sim 15$ events if the disk is maximal. The unbroken logarithmic contours
in the map are almost equally spaced, indicating that microlensing by bulge
deflectors declines roughly exponentially with projected distance from the
Galactic Center. The broken contours are almost horizontal, confirming KP's
insight that the influence of disk lenses is nearly independent of $\ell$.
All this suggests a number of observational tests. First, the disk
lenses quickly become more important than the bulge lenses on the
projected major axis. At the clear window at ($\ell =12.0^\circ,
b=3.0^\circ$), maximal disc models yield $\sim 11$ events per year per
$10^6$ stars, of which more than $70 \%$ are caused by disk lenses. A
large event rate measured at major axis windows with $| \ell | \gta 6^\circ$
is an unambiguous signature of disk deflectors. A second test is to compare
the rate in bulge fields at roughly the same latitude. There are clear
windows at $(\ell =1.0^\circ, b = -3.9^\circ)$ and $(\ell = 5.5^\circ,
b= -3.5^\circ)$, which are candidates for the application of this test.
The contribution of bulge lenses to the rate varies by $\sim 30 \%$, while
the contribution of disk lenses remains unchanged. Variation in the rate
at locations with the same latitude provides evidence that the deflector
population is in the bulge. Inset into Figure 1 are the expected frequencies
of event timescales $t_0$ for the bulge (full lines) and disk (dashed lines)
lenses at two fields currently undergoing intense scrutiny -- one is at
Baade's window, the other is closer to the Galactic Center $(\ell = 2.3^\circ,
b= -2.65^\circ$). As Paczy\'nski (1991) and Griest et al. (1991) point
out, the average timescale is longer for deflectors in the disk. Notice
that this occurs mainly because the dashed distributions have larger tails
towards the longer events. At Baade's Window, the peak timescale is
$\sim 15$ days, irrespective of whether the deflector is a bulge or disk
star. This is longer than that found by KP of $\sim 10$ days. Now, the
efficiency of the microlensing experiments cuts off sharply at short
timescales. If the distribution is peaked at longer timescales, this may
mean that the efficiency has been underestimated, leading to values of
the optical depth derived from the experimental results that are too high.

Figure 2 shows a map for the bar (full lines) and disk (broken lines). The
position angle of the bar's major axis $\theta$ is chosen as $45^\circ$
(c.f. Blitz \& Spergel 1991). The optical depth at Baade's Window is
$5.3 \times 10^{-7}$ for bar lenses, $5.1 \times 10^{-7}$ for disk lenses.
The rate is $\sim 10$ events per year per $10^6$ source stars. The map is
asymmetric. For bar lenses, the number of events is higher at negative
longitudes as compared to positive longitudes. This is because the
near--side of the bar lies in the first Galactic quadrant, and so lines of
sight to detectable stars for $\ell >0$ are on average shorter and pass
through less of the dense inner bar than for $\ell < 0$. This is analogous
to the phenomenon that Crotts (1992) and Gould (1993) noticed for the optical
depth of the inclined disks of M31 and the Large Magellanic Cloud.
For disk lenses, the reverse is true. The rate is slightly higher at positive
longitudes. This is because the rate is averaged over the detectable stars
along the line of sight, with nearer sources carrying more weight than
more distant sources. For $\ell >0$, the sources are on average closer,
thereby amplifying the rate as compared to $\ell <0$. This effect is
model--dependent and diminishes with increasing $\beta$. Udalski et al.
(1994b) are already scanning the windows at $(\ell = -5.0^\circ,
b = -3.5^\circ$) and ($\ell = 5.5^\circ, b = -3.5^\circ$) for possible
evidence of the asymmetric microlensing signal of a bar. In our
model, $\sim 11$ events per year per $10^6$ sources are expected at the
former location, $\sim 7$ at the latter. Now, Blitz \& Spergel's (1991)
estimate of the position angle of the bar's major axis $\theta \sim 45
\pm 20^\circ$ is in broad agreement with Weinberg's (1992) conclusion of
$\theta \sim 36 \pm 10$ based on an analysis of the IRAS 2 micron sources.
But, both results are in apparent conflict with Binney et al. (1991), who
claim $\theta \sim 16\pm 2^\circ$ from an ingenious analysis of the kinematics
of the Galactic Center gas. How does the microlensing rate change as the
viewing angle of the bar is varied? The number of events at Baade's Window
is plotted against $\theta$ and inset into Figure 2. As we view the bar
more nearly down its major axis, the event rate slowly increases to
$\sim 13$ events per year per $10^6$ stars ($\sim 16$ events if the disk
is also maximal). This is a lower limit of the expected amplification. Our
prolate bar (2) looks round when $\theta \sim 0^\circ$ and so does not
properly represent the flattening of the bulge. The neglect of the figure
rotation and internal streaming motions of the bar might also be expected
to lead to an under--estimate of the true rate. Nonetheless, the effect of
distending material along the line of sight seemingly causes only a mild
enhancement. Even if the bar is viewed at $\theta \sim 0^\circ$,
the number of events at Baade's Window is only just greater than for our
oblate axisymmetric model (1). Although it is possible that other bar
models exist with greater amplification, the case for interpreting the high
optical depths reported by Alcock et al. (1994) and Udalski et al. (1994b) as
evidence for a bar remains unproven. Also inset into figure 2 is the
asymmetric signal -- that is, the percentage fractional difference in the
rates at the windows at $(\ell = -5.0^\circ, b= -3.5^\circ)$ and at $(\ell =
5.5^\circ, b = -3.5^\circ)$ -- plotted against viewing angle $\theta$. If the
bar is seen edge--on or pole--on, this asymmetric signal almost vanishes (not
quite, because the windows are not exactly mirror images in $\ell =0$). But,
the asymmetry of the microlensing maps quickly becomes evident as the
orientation of the bar is varied. Even if the bar is viewed at $\sim
16^\circ$, the variation in the rates at $(\ell = -5.0^\circ, b= -3.5^\circ)$
and at $(\ell = 5.5^\circ, b = -3.5^\circ)$ is still large enough to
be detectable.

\section{CONCLUSIONS}

Microlensing maps describe the variation in event rate as a
function of Galactic longitude and latitude. This is valuable in
untangling the contributions of the disk and bulge lenses. We find:

\medskip
\noindent
(1) Oblate axisymmetric models of the bulge yield a total of $\sim
12$ events per year per $10^6$ stars at Baade's Window ($\sim 15$
events if the disk is maximal). Prolate bars can give mild enhancements
of the rate if viewed at $\theta \sim 0^\circ$. Slowly rotating
prolate bars with inclination angles of $\sim 45^\circ$ have lower rates
than oblate models. Unless the bar is seen edge--on or pole--on, the
microlensing maps are asymmetric with larger numbers of events expected
at negative longitudes as compared to positive longitudes. Overall, the
evidence for a bar in the inner Galaxy is strong. So, can the high optical
depths reported in Alcock et al. (1994) and Udalski et al. (1994) be
explained if the microlenses lie in a bar? Our study of a prolate bar
suggests that this is not the whole story, but further modelling
with triaxial and swiftly rotating bars is required for a final answer. The
efficiency of the microlensing experiments depends partly on the
distribution of timescales of events. At Baade's Window, we find this to
peak at $\sim 15$ days -- longer than previous studies (KP). This may mean
the efficiency is being underestimated, accounting for some of the discrepancy
between theoretical and experimental results.

\medskip
\noindent
(2) On the major axis for $|\ell | \gta 6.0^\circ$, the microlensing rate
is dominated by contributions from the disk deflectors. So, measurements
of the rate at major axis windows test directly for disk dark matter or
maximal disk models. For example, maximal disk models provide
$\sim 11$ events per year per $10^6$ sources at the clear window at
($\ell =12.0^\circ, b = 3.0^\circ$), of which over $70 \%$ are caused by
the disk lenses.

\acknowledgments

NWE is particularly indebted to Kim Griest, Kris Stanek and the referee
for numerous helpful comments and suggestions. Useful discussions with
Charles Alcock, Dave Bennett, Kem Cook and Peter Quinn during a visit to
Lawrence Livermore National Laboratory are also acknowledged. NWE is
supported by the Lindemann Trust and the Royal Society.

\begin{figure}
\centerline{FIGURE CAPTIONS}
\caption{Logarithmic contours of equal numbers of microlensing
events for the bulge (1) and disk (3) plotted in the plane of Galactic
longitude $\ell$ and latitude $b$. Unbroken lines depict the contribution from
the bulge lenses (respectively 20, 10, 5 and 2.5 events per year per $10^6$
stars moving outwards from the centre). Broken lines show the contribution
from disk lenses (respectively 5, 2.5 and 1.25 events). The insets show
the distribution of events as a function of timescale $t_0$ for the bulge
(unbroken lines) and disk (broken lines) deflectors at Baade's Window
($\ell = 1.0, b= -3.9^\circ$) and at ($\ell = 2.3^\circ, b = -2.65^\circ$).}

\caption{Logarithmic contours of equal numbers of microlensing events
for the bar (2) and disk (3). The position angle of the major axis of the
bar $\theta$ is $45^\circ$. The unbroken lines are the contributions from
the bar lenses, the broken lines are the contributions from the disk
lenses. The inset (a) shows the variation of the microlensing rate at
Baade's Window (measured in events per year per $10^6$ source stars)
with the viewing angle of the bar $\theta$. The inset (b) shows the
variation of the asymmetric signal -- defined as the percentage
fractional difference in the total rate at the clear windows at
$(\ell = -5.0^\circ, b = -3.5^\circ)$ and $(\ell = 5.5^\circ, b =
-3.5^\circ$) -- with viewing angle $\theta$. In inset (b), we have
not separated the contributions from bar and disk lenses. Of course,
the contribution to the asymmetric signal from the disk lenses is
almost negligible.}
\end{figure}

\end{document}